\documentclass[conference]{IEEEtran}
\IEEEoverridecommandlockouts
\usepackage{cite}
\usepackage{amsmath,amssymb,amsfonts}
\usepackage{algorithmic}
\usepackage{textcomp}
\usepackage{xcolor}
\usepackage[pdftex]{graphicx}
\usepackage{epstopdf}

\usepackage{amsmath,amssymb,enumerate} 
\usepackage{bm} 
\usepackage{indentfirst} 
\usepackage{empheq} 
\usepackage{here} 
\newcommand{\dd}{\mathrm{d}}

\newcommand{\ii}{\mathrm{i}}

\def\BibTeX{{\rm B\kern-.05em{\sc i\kern-.025em b}\kern-.08em
    T\kern-.1667em\lower.7ex\hbox{E}\kern-.125emX}}
\begin{document}

\title{Modeling of Online Echo-Chamber Effect
Based on the Concept of Spontaneous Symmetry Breaking}

\author{\IEEEauthorblockN{Masaki Aida}
    \IEEEauthorblockA{
        \textit{Tokyo Metropolitan University} \\
Tokyo 191--0065, Japan \\
aida@tmu.ac.jp}
\and
\IEEEauthorblockN{Ayako Hashizume}
    \IEEEauthorblockA{
        \textit{Hosei University} \\
Tokyo 194--0298, Japan \\
hashiaya@hosei.ac.jp}
}

\maketitle

\begin{abstract}
The online echo-chamber effect is a phenomenon in which beliefs that are far from common sense are strengthened within relatively small communities formed within online social networks.
Since it is significantly degrading social activities in the real world, we should understand how the echo-chamber effect arises in an engineering framework to realize countermeasure technologies.
This paper proposes a model of the online echo-chamber effect by introducing the concept of spontaneous symmetry breaking to the oscillation model framework used for describing online user dynamics.
\end{abstract}


\section{Introduction}
With the development of information networks, user activities on online social networks (OSNs) using social networking services (SNSs) are becoming more active.
While OSNs can streamline people's activities in the real world, they can also cause social problems such as the online flaming phenomenon and online echo-chamber effect.
Therefore, understanding user dynamics in OSNs is a crucial issue.

The oscillation model is known to be able to describe user dynamics in OSNs~\cite {Aida2018,Aida2016}.
This is a minimal model that applies the wave equation on networks.
In the oscillation model concept, user activities influence each other through the OSN, and the wave equation models the propagation of their influence through the OSN at a finite speed.
Since the oscillation model well models the online flaming phenomenon, this paper focuses on modeling the online echo-chamber effect.

\begin{figure}[t]
\begin{center}
\includegraphics[width=0.9\linewidth]{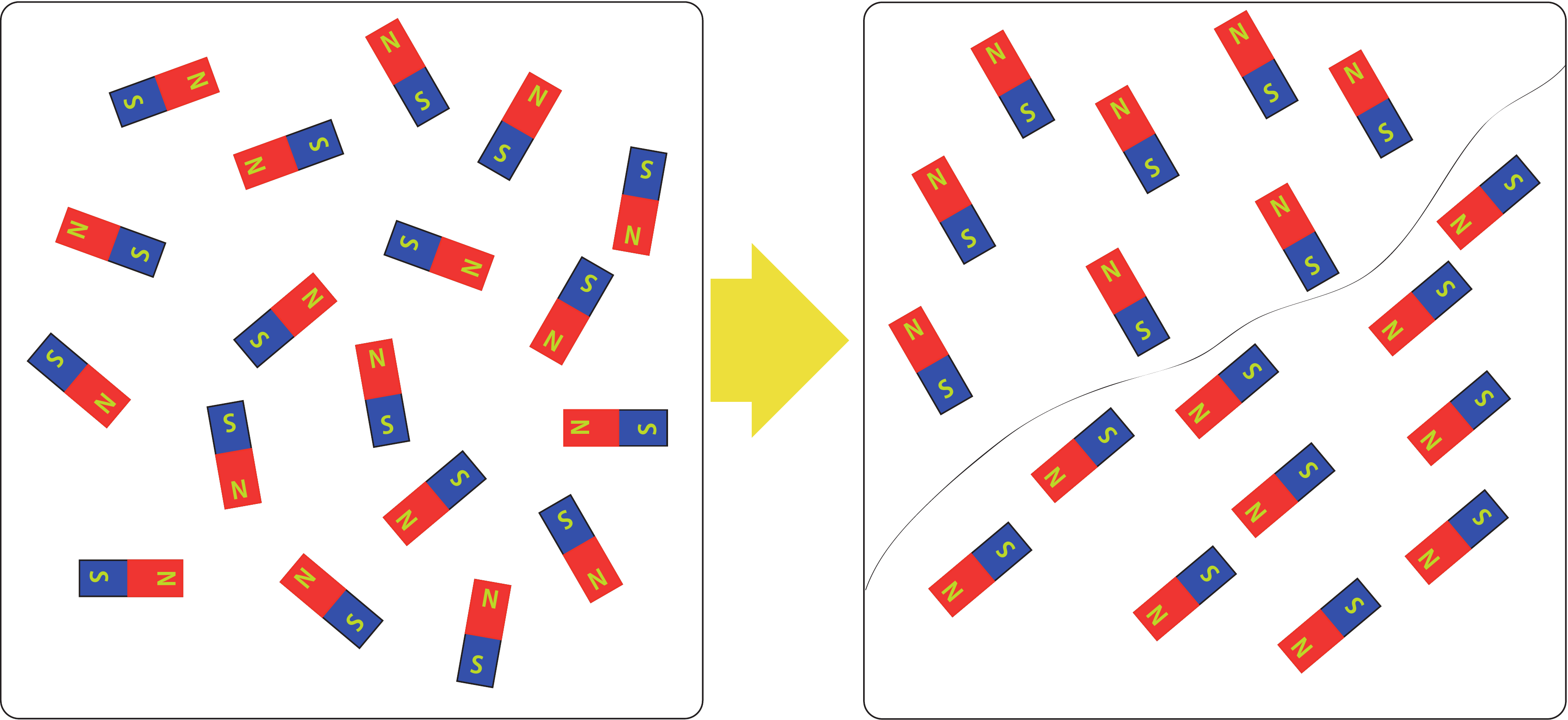}
\caption{Image of spontaneous magnetization of ferromagnetic material}
\label{fig:mag}
\end{center}
\end{figure}

The online echo-chamber effect is a phenomenon in which beliefs that are far from common sense are strengthened within relatively small communities that develop in OSNs.
This begins with forming a closed cluster of users, with whom they become aligned on a particular opinion.
Such a phenomenon is similar to the concept of spontaneous symmetry breaking (SSB) \cite{Galam}.
SSB refers to a situation in which an asymmetric state emerges that is stable due to the symmetry originally possessed by the system breaking spontaneously.
SSB is a theoretical model that has been used to describe spontaneous magnetization (see Fig.~\ref{fig:mag}).
The atoms of ferromagnetic material are small magnets, and no overall magnetism is exhibited if each atom is oriented in a random direction. 
However, if the atoms' directions become aligned for some reason, the material acts as a single large magnet. 

%
\begin{figure}[b]
\begin{center}
\begin{tabular}{cc}
\includegraphics[width=0.42\linewidth]{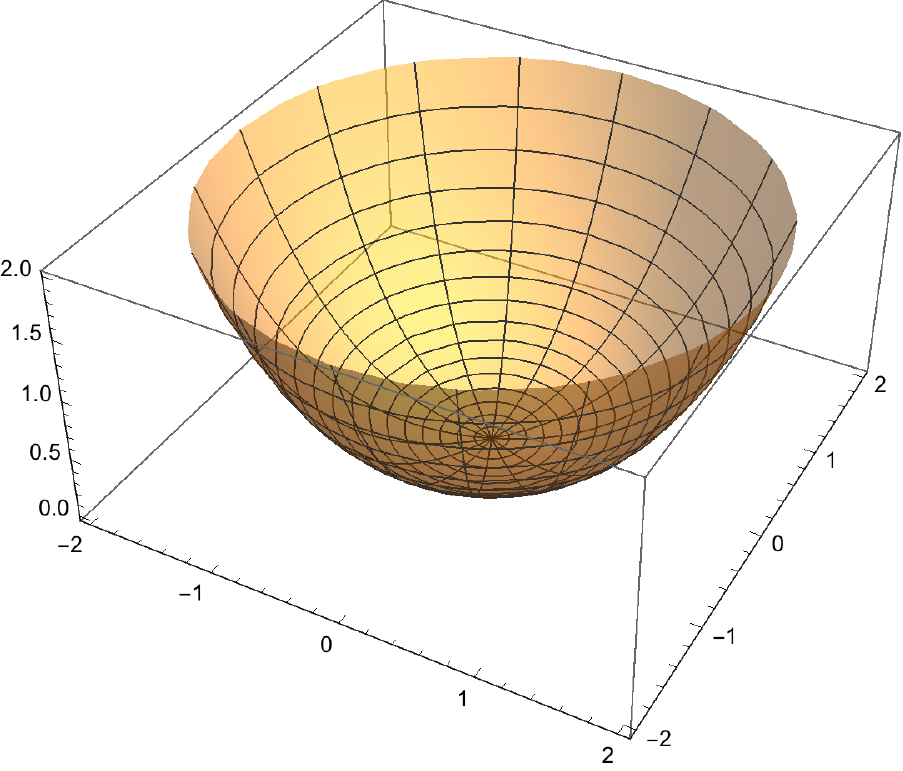} 
&\includegraphics[width=0.42\linewidth]{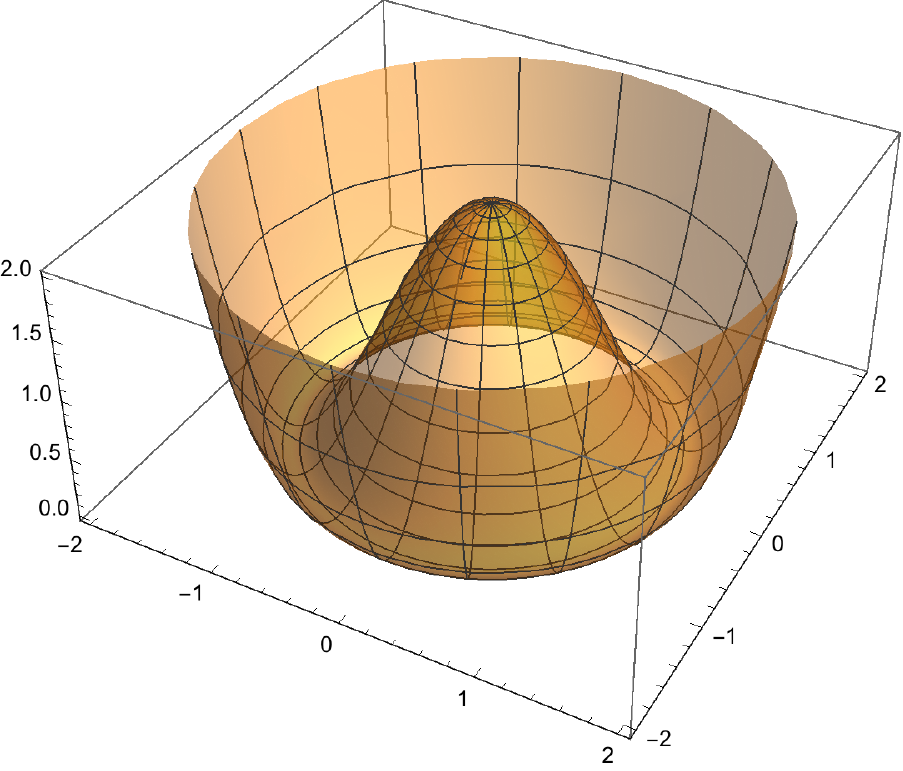}\\
{\small (a) Before symmetry breaking}  
& {\small (b) After symmetry breaking}
\end{tabular}
\caption{The shapes of the potential function before and after SSB}
\label{fig:SSB}
\end{center}
\end{figure}

Since SSB is essentially a theoretical model of quantum theory, there is a problem in applying SSB as is to modeling social phenomena.
Figure \ref{fig:SSB} shows the changes in the potential function before and after SSB.
Before SSB, it is stable at the origin. 
When SSB occurs, it changes to the shape in the right called the Mexican hat potential. 
The origin becomes unstable and stabilizes at the valley bottom in an arbitrarily selected direction.
This corresponds to the phenomenon that the ferromagnetic atoms spontaneously align in a specific direction.
After SSB, new dynamics that orbit the valley bottom, which did not exist before SSB, can now exist; this is called the Nambu-Goldstone mode (N-G mode).
The superficial analogy provides no understanding of how the N-G mode corresponds to the attributes of actual OSNs.

Since the fundamental equation of user dynamics derived from the oscillation model has a structure similar to the Dirac equation of relativistic quantum mechanics, SSB can be expected to be naturally incorporated into the framework of the oscillation model.
However, it is very artificial to introduce the Mexican hat potential function {\em a priori}; it is necessary to identify the OSN structural change that works in the same way as the Mexican hat potential.
Also, it is necessary to elucidate the emergence of the N-G mode in OSNs.
This paper discusses a model of the online echo-chamber effect that applies the concept of SSB to the framework of the oscillation model.

\section{Oscillation Model for User Dynamics in OSNs}
\label{sec:OM}
This section briefly summarizes the fundamental equations of the oscillation model.
First, for nodes $i,\,j \in V$ of directed graph $G (V, E)$, which represents the structure of an OSN with $n$ users, if the weight of directed link $(i \rightarrow j) \in E$ is given as $w_{ij}$, the adjacent matrix $\bm{\mathcal{A}} = [\mathcal{A}_{ij}]_{1\le i,j \le n}$ is defined as 
\begin{align}
\mathcal{A}_{ij} := \left\{
\begin{array}{cl}
w_{ij},&  \quad (i\rightarrow j) \in E,\\
0,& \quad (i\rightarrow j) \not\in E.
\end{array}
\right. 
\end{align}
Also, given nodal (weighted) out-degree $d_i := \sum_{j\in \partial i} w_{ij}$, 
the degree matrix is defined as 
\begin{align}
\bm{\mathcal{D}} := \mathrm{diag}(d_1,\,\dots\,d_n). 
\end{align}
Here, $\partial i$ denotes the set of adjacent nodes of out-links from node $i$. 
Next, the Laplacian matrix of the directed graph representing the OSN structure is defined by 
\begin{align}
\bm{\mathcal{L}} := \bm{\mathcal{D}} - \bm{\mathcal{A}}.
\end{align}
To eliminate situations that trigger online flaming phenomena, we assume that all eigenvalues of $\bm{\mathcal{L}}$ are real.

Let the state vector of users at time $t$ be 
\[
\bm{x}(t) := {}^t\!(x_1(t),\,\dots,\,x_n(t)), 
\]
where 
$x_i(t)$ $(i=1,\,\dots,\,n)$ denotes the user state of node $i$ at time $t$. 
Then, the wave equation for the OSN is written as 
\begin{align}
\frac{\dd^2}{\dd t^2} \, \bm{x}(t) = - \bm{\mathcal{L}} \, \bm{x}(t). 
\label{eq:wave-eq}
\end{align}
Here, in addition to simply finding the solution $\bm{x}(t)$ of the wave equation (\ref{eq:wave-eq}), it is desirable to be able to describe what kind of OSN structure impacts user dynamics. 
In other words, we want to describe the causal relationship between OSN structure and user dynamics. 
To achieve this, we need to develop a first-order differential equation with respect to time (hereinafter referred to as the fundamental equation)~\cite{Aida-book,Aida2020}.

\begin{figure*}[t]
\begin{center}
\begin{tabular}{ccccc}
\includegraphics[width=0.255\linewidth]{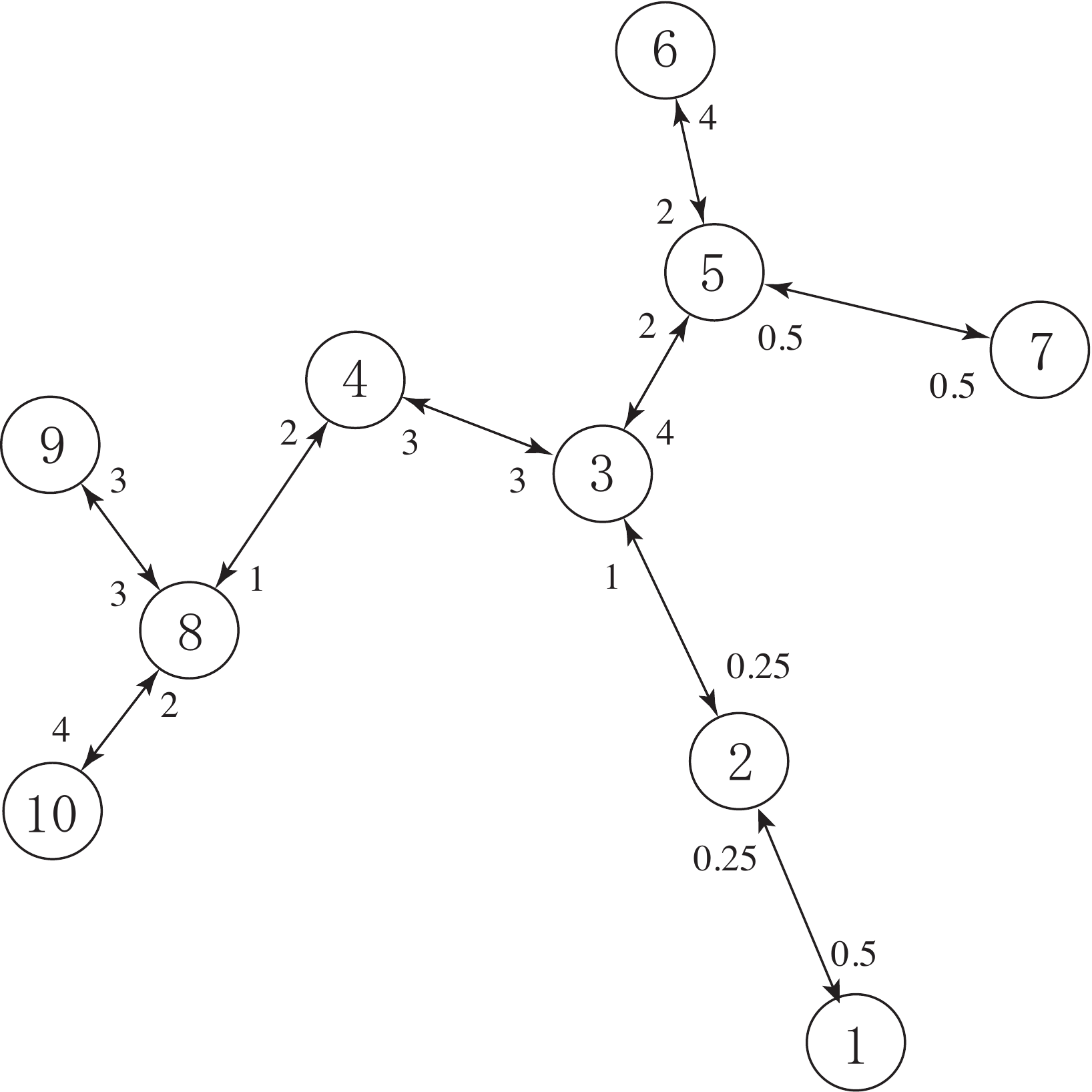} & \quad & 
\includegraphics[width=0.31\linewidth]{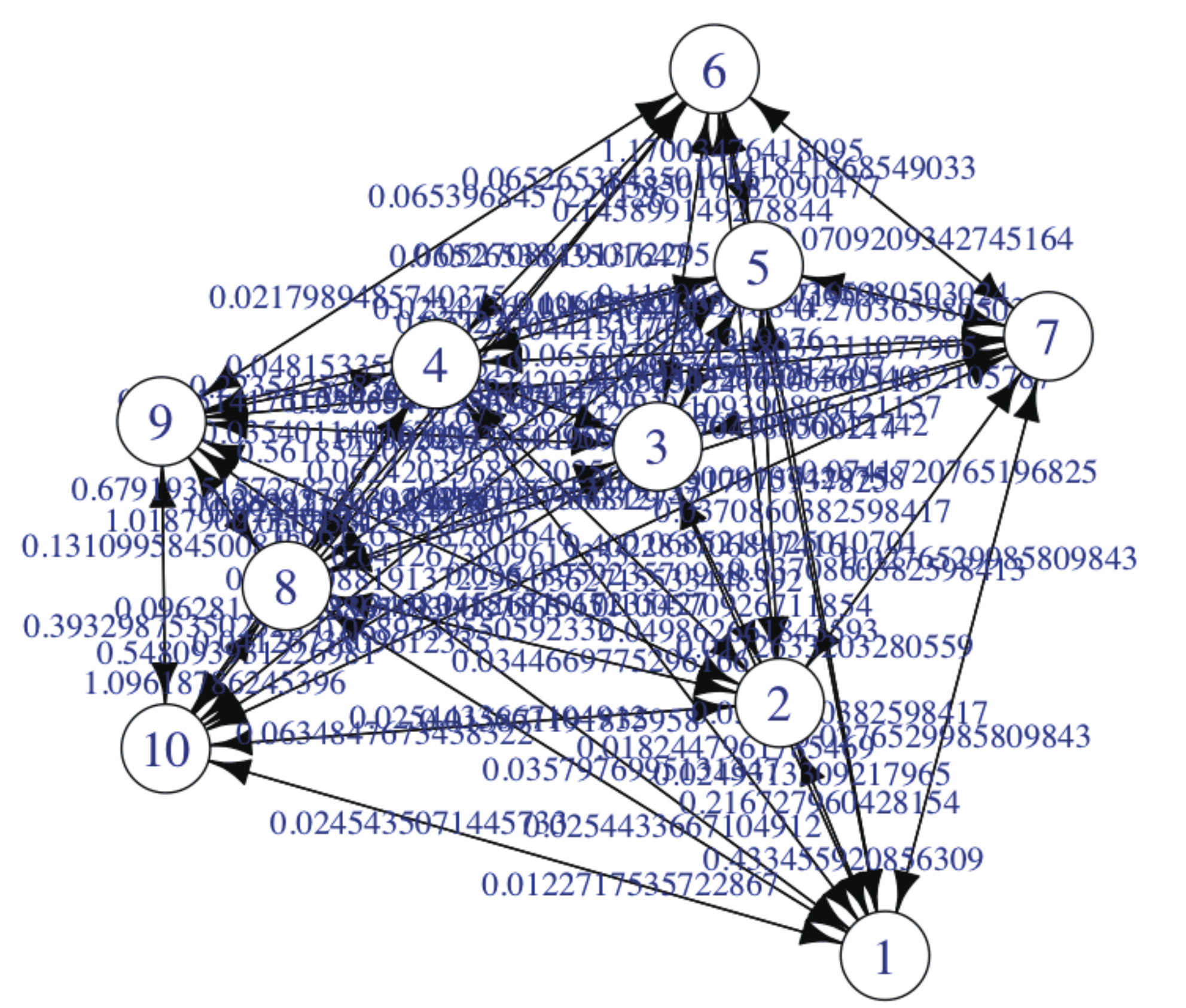} & \quad & 
\includegraphics[width=0.255\linewidth]{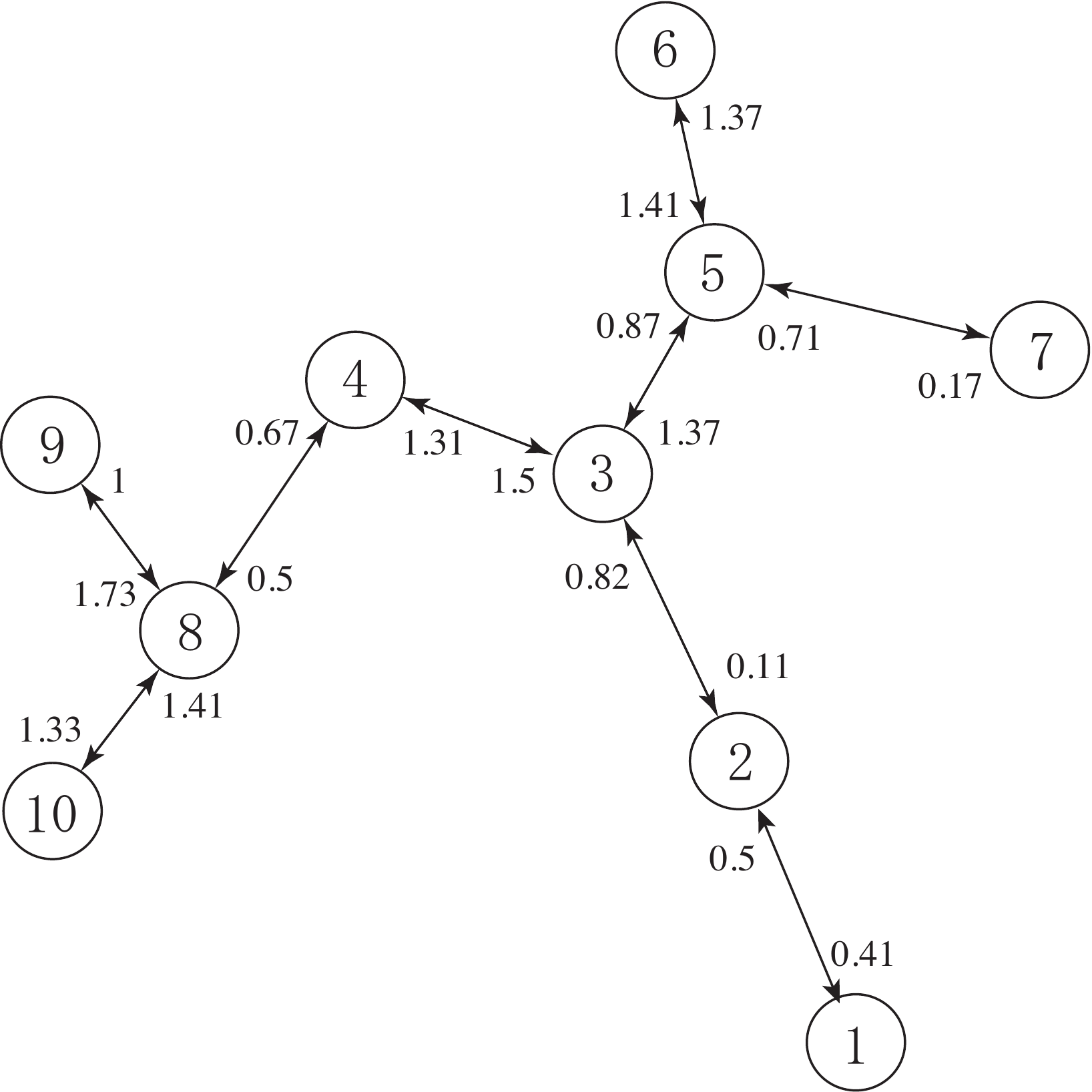}\\
Laplacian matrix $\bm{\mathcal{L}}$ && square root matrix $\sqrt{\bm{\mathcal{L}}}$ && semi-normalized Laplacian matrix $\bm{\mathcal{H}}$
\end{tabular}
\caption{An example of network structures represented by the Laplacian matrix $\bm{\mathcal{L}}$, 
the square root matrix $\sqrt{\bm{\mathcal{L}}}$, and the semi-normalized Laplacian matrix $\bm{\mathcal{H}}$}
\label{fig:1}
\end{center}
\end{figure*}

Not only is the fundamental equation a first-order differential equation, but its solution must also be the solution of the wave equation (\ref{eq:wave-eq}).
The simplest way to realize this is as follows. 
By using the positive semi-definite matrix $\sqrt{\bm{\mathcal{L}}}$, which is the square root of $\bm{\mathcal {L}}$, we introduce the following fundamental equation (double-sign corresponds), 
\begin{align}
\pm\ii \,\frac{\dd}{\dd t}\, \bm{x}^\pm(t) &= \sqrt{\bm{\mathcal{L}}} \, \bm{x}^\pm(t),
\label{eq:fundamental-0}
\end{align}
where $\sqrt{\bm{\mathcal{L}}}$ is uniquely determined for $\bm{\mathcal {L}}$. 
The two fundamental equations (\ref{eq:fundamental-0}) can, by using a $2n$-dimensional state vector, be expressed as a single expression of 
\begin{align}
\ii \,\frac{\dd}{\dd t}\, \bm{\Hat{x}}(t) = \left(\sqrt{\bm{\mathcal{L}}} \otimes 
\begin{bmatrix}
+1 & 0\\
0 & -1
\end{bmatrix}
\right) \bm{\Hat{x}}(t),
\label{eq:fundamental-1}
\end{align}
where for $\bm{x}^\pm(t) = {}^t\!(x_1^\pm(t),\,\dots,\,x_n^\pm(t))$ (double-sign corresponds), 
the $2n$-dimensional state vector $\bm{\Hat{x}}(t)$ is defined as 
\[
\bm{\Hat{x}}(t) := \bm{x}^+(t) \otimes 
\begin{pmatrix}
1\\
0
\end{pmatrix}
+ \bm{x}^-(t) \otimes 
\begin{pmatrix}
0\\
1
\end{pmatrix}. 
\]
Also, $\otimes$ denotes the Kronecker product~\cite{brewer}. 
We call (\ref{eq:fundamental-1}) the boson-type fundamental equation. 

The boson-type fundamental equation~(\ref{eq:fundamental-1}) has a technical issue.
Even if the Laplacian matrix $\bm{\mathcal {L}}$ is sparse, its square root matrix $\sqrt{\bm{\mathcal{L}}}$ is generally a complete graph (see Fig.~\ref{fig:1}).
In ordinary large-scale social networks, it is implausible to assume a situation in which all users are directly connected. 
The matrix that appears in the fundamental equation must completely reflect the OSN link structure (whether there is a link between OSN nodes). 
Therefore, the solutions of the boson-type fundamental equation~(\ref{eq:fundamental-1}) cannot exist in OSNs that have structures other than complete graphs.

To avoid this problem, we introduce another fundamental equation. 
Let us introduce a new matrix, $\bm{\mathcal{H}}$, as follows. 
\begin{align}
\bm{\mathcal{H}} := \sqrt{\bm{\mathcal{D}}^{-1}} \, \bm{\mathcal{L}} = \sqrt{\bm{\mathcal{D}}} - \sqrt{\bm{\mathcal{D}}^{-1}} \, \bm{\mathcal{A}}, 
\label{eq:def_H}
\end{align}
where $\sqrt{\bm{\mathcal{D}}} := \text{diag}(\sqrt{d_1},\,\dots,\,\sqrt{d_n})$. 
As is well-known, the normalized Laplacian matrix is defined as
\[
\bm{\mathcal{N}} := \sqrt{\bm{\mathcal{D}}^{-1}}\, \bm{\mathcal{L}} \,  \sqrt{\bm{\mathcal{D}}^{-1}} = \bm{I} -   \sqrt{\bm{\mathcal{D}}^{-1}} \, \bm{\mathcal{A}}\, \sqrt{\bm{\mathcal{D}}^{-1}}; 
\]
so we call $\bm{\mathcal{H}}$ the semi-normalized Laplacian matrix. 
Here, $\bm{I}$ is the $n\times n$ unit matrix. 
The semi-normalized Laplacian matrix $\bm{\mathcal{H}}$ is also a Laplacian matrix of a certain directed graph, which has the same link structure as $\bm{\mathcal{L}}$ as regards link existence and absence (see Fig.~\ref{fig:1}).

Using the semi-normalized Laplacian matrix $\bm{\mathcal{H}}$, a new fundamental equation for user dynamics can be written as follows:
\begin{align}
  \ii \, \frac{\dd}{\dd t} \,  \bm{\hat{x}}(t)
  = \left(\bm{\mathcal{H}} \otimes \bm{\hat{a}}
 + \bm{\sqrt{\mathcal{D}}} \otimes \bm{\hat{b}}
 \right)\bm{\hat{x}}(t), 
 \label{eq:fundamental1}
\end{align}
where $\bm{\hat{a}}$ and $\bm{\hat{b}}$ are $2\times 2$ matrices defined as 
\[
\bm{\hat{a}} = 
\frac{1}{2}
\begin{bmatrix}
 +1 & +1\\ 
 -1 & -1 
\end{bmatrix}, \quad 
\bm{\hat{b}} = 
\frac{1}{2}  
\begin{bmatrix}
 +1 & -1\\ 
 +1 & -1 
 \end{bmatrix} ,
\]
and $\bm{\hat{x}}(t)$ is the $2n$-dimensional state vector. 
Solution $\bm{x}(t)$ of the original wave equation (\ref{eq:wave-eq}) can be obtained from $\bm{\hat{x}}(t)$ by 
\begin{align}
\bm{x}(t) = (\bm{I}\otimes (1,1))\,\bm{\hat{x}(t)}. 
\label{eq:x}
\end{align}

The fundamental equation is similar to the Dirac equation of relativistic quantum mechanics, and its feature is that $\bm{\hat{a}}$ and $\bm{\hat{b}}$ satisfy the anti-commutation relation: 
\begin{align}
\{\bm{\hat{a}},\bm{\hat{b}}\} := \bm{\hat{a}}\,\bm{\hat{b}} + \bm{\hat{b}}\,\bm{\hat{a}} = \bm{\hat{e}}, \quad 
\bm{\hat{a}}^2 = \bm{\hat{b}}^2 = \bm{\hat{\mathrm{o}}},
\label{eq:anti-comm}
\end{align}
where $\bm{\hat{e}}$ denotes the $2\times 2$ unit matrix and $\bm{\hat{\mathrm{o}}}$ denotes the null matrix. 
Therefore, we call (\ref{eq:fundamental1}) the fermion-type fundamental equation. 

\section{Echo-Chamber Phenomena}
\label{sec:E-C}
\begin{figure*}[t]
\begin{center}
\includegraphics[width=0.85\linewidth]{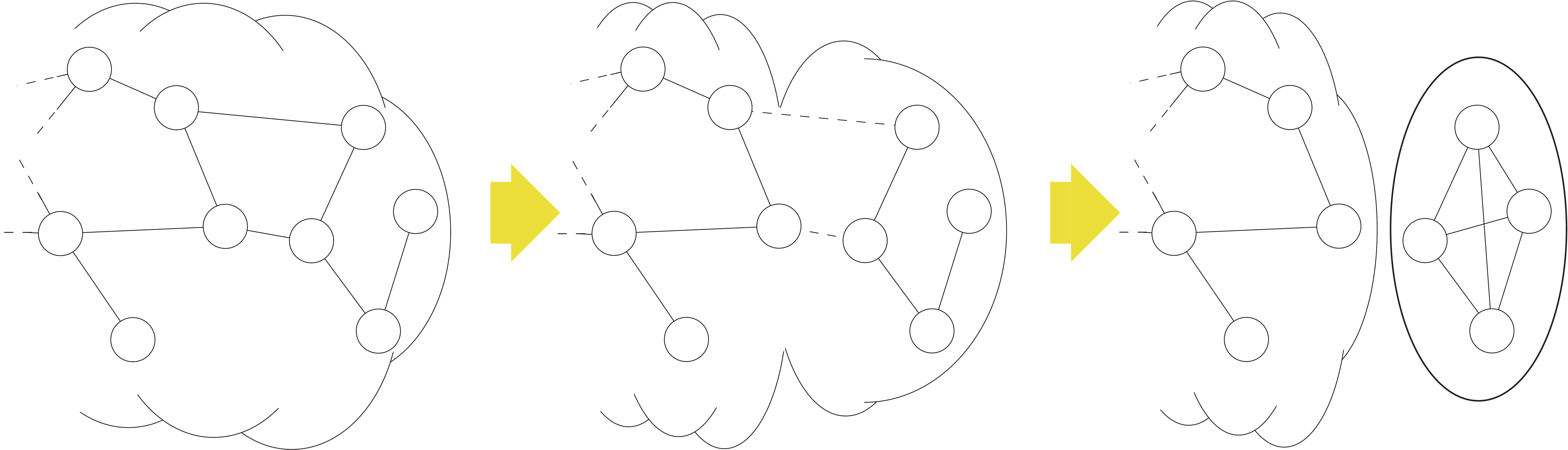}
\caption{Structural change in OSNs that yields an isolated cluster}
\label{fig:4}
\end{center}
\end{figure*}

One model proposed to explain the occurrence of SSB posits that a relatively small subnetwork corresponding to a closed community detaches itself from its underlying OSN, and the isolated subnetwork becomes a complete graph \cite{ITC32} (see Fig.~\ref{fig:4}).

The isolated subnetworks can oscillate around a different equilibrium point than the OSN before isolation.
Therefore, a biased state is realized.
In addition to the fermion-type fundamental equation's solution, a new solution of the boson-type fundamental equation~(\ref{eq:fundamental-1}) can exist because the isolated subnetwork is in effect a complete graph. 
These two features can be associated with SSB features, and the latter can be interpreted as the N-B mode.

Here, based on the fermion-type fundamental equation (\ref{eq:fundamental1}), we show the simplest case. 
Let us consider the situation in which the weights of all isolated subnetwork links have the same value, $w$.
This is a situation in which the link's weight, which indicates the strength of the relationship between users, has increased to the limit and is saturated because the discussion in the isolated community is fully activated.
At this time, if the number of users in the isolated community is $n$, the nodal degree is the same for all nodes, and $d = (n-1)\,w$.
Also, in the corresponding Laplacian matrix, all eigenvalues other than $0$ are duplicated, and the eigenvalues are denoted as $\lambda = \omega^2 = n\,w$.

In this situation, since the degree matrix is $\bm{\mathcal{D}} = d\,\bm{I}$, the matrices $\bm{\mathcal{H}}$ and $\sqrt{\bm{\mathcal{D}}}$ can be diagonalized simultaneously. 
Therefore, the fermion-type fundamental equation (\ref{eq:fundamental1}) can be transformed so that both the matrices $\bm{\mathcal{H}}$ and $\sqrt{\bm{\mathcal{D}}}$ are diagonalized.
Thus, the fermion-type fundamental equation is expressed in the block diagonalized form of $2\times 2$. 
The transformed equations for all the eigenvalues other than the eigenvalue of $0$ of the Laplacian matrix, are the same and can be written as follows:
\begin{align}
\ii \, \frac{\dd}{\dd t} \bm{\psi}(t) 
&= \left(\frac{\omega^2}{2\,\sqrt{d}} 
  \begin{bmatrix}
   +1 & +1\\ 
   -1 & -1
  \end{bmatrix}
  + \frac{\sqrt{d}}{2} 
  \begin{bmatrix}
   +1 & -1\\ 
   +1 & -1
  \end{bmatrix}
  \right) \bm{\psi}(t)
\notag\\
&=
  \begin{bmatrix}
   +\frac{\omega^2+d}{2\,\sqrt{d}} & +\frac{\omega^2-d}{2\,\sqrt{d}}\\ 
   -\frac{\omega^2-d}{2\,\sqrt{d}} & -\frac{\omega^2+d}{2\,\sqrt{d}}
  \end{bmatrix}
\, \bm{\psi}(t).  
\label{eq:block-diag1}
\end{align}
Here, if the two-dimensional vector of the solution of (\ref{eq:block-diag1}) is denoted as  
\begin{align}
\bm \psi (t) = 
 \begin{pmatrix}
  \psi^+ (t) \\ 
  \psi^- (t)
 \end{pmatrix},
\end{align}
and set the Ansatz of (\ref{eq:block-diag1}) as 
\begin{align}
  \psi^\pm(t) := \exp\!\left(\mp \ii\theta^\pm (t)\right),
  \label{eq:ansatz}
\end{align}
in double sign correspondence; this yields
\begin{align}
\frac{\dd}{\dd t} \theta^\pm (t) = \frac{\omega^2 + d}{2 \sqrt{d}} + \frac{\omega^2 - d}{2 \sqrt{d}}\exp\!\left(\pm \mathrm{i}\left(\theta^+(t) + \theta^-(t)\right)\right). 
\label{eq:d_theta/dt_1}
\end{align}
Since $\theta^\pm(t)$ is a complex number in general, by substituting 
\[
\theta^\pm(t) = \mathrm{Re}[\theta^\pm(t)] + \ii \, \mathrm{Im}[\theta^\pm(t)]
\]
into (\ref{eq:d_theta/dt_1}), we obtain the temporal evolutions of the real and the imaginary parts of $\theta^\pm(t)$ as 
\begin{align}
&\frac{\dd}{\dd t} \, \mathrm{Re}[\theta^\pm(t)] 
\notag\\
&= \frac{\omega^2+d}{2\sqrt{d}} + C^\pm(t) \, \cos\!\left(\mathrm{Re}[\theta^+ (t)] + \mathrm{Re}[\theta^- (t)]\right) 
\notag\\
&= \frac{\omega^2+d}{2\sqrt{d}} 
\notag\\
&\qquad {}+ C^\pm(t) \, \sin\!\left(-\left(\mathrm{Re}[\theta^\mp (t)]-\frac{\pi}{2}\right) - \mathrm{Re}[\theta^\pm (t)]\right), 
\label{eq:real1}\\
&\frac{\dd}{\dd t} \, \mathrm{Im}[\theta^\pm(t)] 
= \pm C^\pm(t) \, \sin\!\left(\mathrm{Re}[\theta^+ (t)] + \mathrm{Re}[\theta^- (t)]\right),  
\label{eq:imaginary1}
\end{align}
where 
\begin{align}
C^\pm(t) := \frac{\omega^2 - d}{2 \sqrt{d}}\,\exp\!\left(\mp \left(\mathrm{Im}[\theta^+(t)] + \mathrm{Im}[\theta^-(t)]\right)\right). 
\label{eq:C}
\end{align}

From this result, the following properties of the solution can be predicted.
Note that the temporal evolution (\ref{eq:real1}) of the real part of $\theta^\pm (t)$ has a structure similar to that of the Kuramoto model \cite{Kuramoto}, since $C^\pm(t) > 0$.
The difference from the Kuramoto model is that $C^\pm(t)$ is not a constant. 
Now, if $C^\pm(t)$ is large enough and phase synchronization occurs as in a Kuramoto model, the following states are realized: 
\[
\mathrm{Re}[\theta^+ (t)] + \mathrm{Re}[\theta^- (t)] = +\frac{\pi}{2}. 
\]
Along with this, the temporal change (\ref{eq:imaginary1}) of the imaginary part of $\theta^\pm (t)$ becomes
\begin{align}
\frac{\dd}{\dd t} \, \mathrm{Im}[\theta^\pm(t)] 
&= \pm C^\pm(t), 
\label{eq:result1}
\end{align}
because the $\sin$ function part of (\ref{eq:imaginary1}) becomes $+1$.
Therefore, $\theta^+(t)$ increases and $\theta^-(t)$ decreases with time.
In both cases, these changes increase the amplitude of $\psi^\pm$ according to the Ansatz (\ref{eq:ansatz}).
That is since the amplitude of the Ansatz (\ref{eq:ansatz}) is denoted as 
\[
|\psi^\pm(t)| = \exp\!\left(\pm\mathrm{Im}[\theta^\pm(t)]\right),
\]
the amplitude $|\psi^+(t)|$ increases if $\mathrm{Im}[\theta^+(t)]$ increases and the amplitude $|\psi^-(t)|$ increases if $\mathrm{Im}[\theta^-(t)]$ decreases. 
This result means that user activity in the isolated community is activated, and it is considered that this describes the occurrence of the online echo-chamber effect.

\section{Conclusion}
\label{sec:conclusion}
This paper has proposed a model that introduces the concept of SSB to explain the online echo chamber phenomenon.
SSB originated as a theoretical model in quantum mechanics to explain the spontaneous magnetization of ferromagnetism. 
Still, an analogy with a phenomenon in which opinions are aligned in social communities has also been discussed.
However, superficial analogies failed to identify the correspondence between social phenomenon and the Nambu-Goldstone mode, new dynamics that appear when SSB occurs.
This paper has shown how the OSN structural changes that reflect the user dynamics of OSN correspond to spontaneous symmetry breaking in the framework of the oscillation model; we can now explain the emergence of new dynamics corresponding to the Nambu-Goldstone mode.
This paper focused on the simplest case of user dynamics for an isolated community based on this framework. 
It showed the possibility of a phenomenon in which the strength of user dynamics increases well mirrors the online echo-chamber effect. 
\section*{Acknowledgement}
This research was supported by Grant-in-Aid for Scientific Research 19H04096 and 20H04179 from the Japan Society for the Promotion of Science (JSPS).



\end{document}